\pdfoutput=1

\documentclass[letterpaper,twocolumn,10pt]{article}
\usepackage{usenix}

\usepackage[numbers]{natbib}
\usepackage{svg}

\usepackage{hyperref}
\usepackage{xcolor}
\hypersetup{
    colorlinks,
    linkcolor={red!50!black},
    citecolor={blue!50!black},
    urlcolor={blue!80!black}
}

\usepackage{microtype}
\usepackage{graphicx}

\usepackage[caption=false,font=footnotesize,labelfont=sf,textfont=sf]{subfig}

\usepackage[final]{changes}
\usepackage{amsfonts}
\usepackage{amsmath}
\usepackage[capitalise]{cleveref}
\usepackage{wrapfig}
\usepackage{booktabs} %
\usepackage{longtable}
\usepackage{verbatim}
\usepackage{svg}
\usepackage{makecell}
\usepackage{hyperref}
\usepackage{multirow}
\usepackage{enumitem}
\usepackage{mathtools}

\definecolor{myaddedcolor}{rgb}{0.1, 0.7, 0.3} %
\setaddedmarkup{\textcolor{myaddedcolor}{#1}}

\newcommand{\attribute}{\texttt{X}}

\newcommand{\change}[1]{#1}

\makeatletter
\renewcommand\paragraph{\@startsection{paragraph}{4}{\z@}%
{1ex \@plus.2ex \@minus.2ex}%
{-0.5em}%
{\normalfont\normalsize\bfseries}}
\makeatother
\usepackage{tikz}

\let\oldparagraph=\paragraph
\renewcommand\paragraph[1]{\oldparagraph{#1.}}

\title{Privacy Side Channels in Machine Learning Systems}

\author{Edoardo Debenedetti$^1$ \qquad Giorgio Severi$^2$ \qquad Nicholas Carlini$^3$ \qquad Christopher A. Choquette-Choo$^3$\\  Matthew Jagielski$^3$ \qquad Milad Nasr$^3$ \qquad
Eric Wallace$^4$ \qquad Florian Tramèr$^1$ \\
\emph{$^1$ETH Zurich \qquad $^2$Northeastern University \qquad $^3$Google DeepMind \qquad$^4$UC Berkeley}}

\begin{document}

\pagenumbering{gobble}

\maketitle

\begin{abstract}

Most current approaches for protecting privacy in machine learning (ML) assume that models exist in a vacuum. Yet, in reality, these models are part of larger \textit{systems} that include components for training data filtering, output monitoring, and more.
In this work, we introduce \emph{privacy side channels}: attacks that exploit these system-level components to extract private information at far higher rates than is otherwise possible for standalone models.
We propose four categories of side channels that span the entire ML lifecycle (training data filtering, input preprocessing, output post-processing, and query filtering) and allow for enhanced membership inference, data extraction, and even novel threats such as extraction of users' test queries.
For example, we show that deduplicating training data before applying differentially-private training creates a side-channel that completely invalidates any provable privacy guarantees.
We further show that systems which block language models from regenerating training data can be exploited to exfiltrate private keys contained in the training set---even if the model did not memorize these keys. 
Taken together, our results demonstrate the need for a holistic, end-to-end privacy analysis of machine learning systems.
\end{abstract}

\section{Introduction}
In the absence of safeguards, machine learning (ML) models leak private information about their training data~\citep{fredrikson2015model,shokri2017membership}. 
Numerous methods have been proposed to measure and mitigate privacy leakage, including formal techniques~\citep{dwork2006calibrating,abadi2016deep,choquette2022multi} and heuristics~\citep{lee2021deduplicating,kandpal2022deduplicating}.
However, existing \added{privacy-preserving} methods largely assume that ML models \emph{exist in a vacuum}, when in reality ML models are part of larger \textit{systems} that include components for training data filtering, input preprocessing, output monitoring, and more. These  system-level components are widely incorporated into real-world ML systems to maximize accuracy, security, and robustness.

In this work, we introduce \emph{privacy side channels}: attacks that exploit system-level components to extract private information at much higher rates than is otherwise possible for isolated ML models. 
We show that adaptive adversaries of varying strengths---ranging from black-box query access to data poisoning capabilities---can mount privacy attacks that are otherwise impossible without side channels (e.g., revealing test inputs). \change{We evaluate the impact of these side-channel attacks on end-to-end systems, including GitHub Copilot---a black-box production system with millions of users.}
Concretely, we propose four categories of attacks that span the entire ML lifecycle (overview in Figure~\ref{fig:main}):

\begin{itemize}[topsep=5pt, itemsep=5pt, leftmargin=10pt]
\item \textbf{Training data filtering (Section~\ref{sec:train_filters}).}  
Most large-scale training sets are filtered to remove duplicates and abnormal examples~\citep{lee2021deduplicating, kandpal2022deduplicating, anil2023palm}. 
We demonstrate that data filters introduce side channels because they create dependencies between different users' data.
In turn, adversaries can amplify privacy attacks by inserting poison examples that maximize these dependencies.
Perhaps most surprisingly, we show that data deduplication~\citep{lee2021deduplicating}---a technique designed to \emph{improve} privacy---can make privacy \emph{worse}, even causing violations of naive differential privacy (DP) guarantees (Section~\ref{sec:dp}). Aside from deduplication, we propose similar attacks for defenses against data poisoning~\citep{fung2018mitigating, chen2018detecting, carlini2021poisoning}.

\begin{figure*}
    \centering
    \includegraphics[trim={0.0cm, 6.4cm, 0.5cm, 0.0cm}, clip,width=0.9\linewidth]{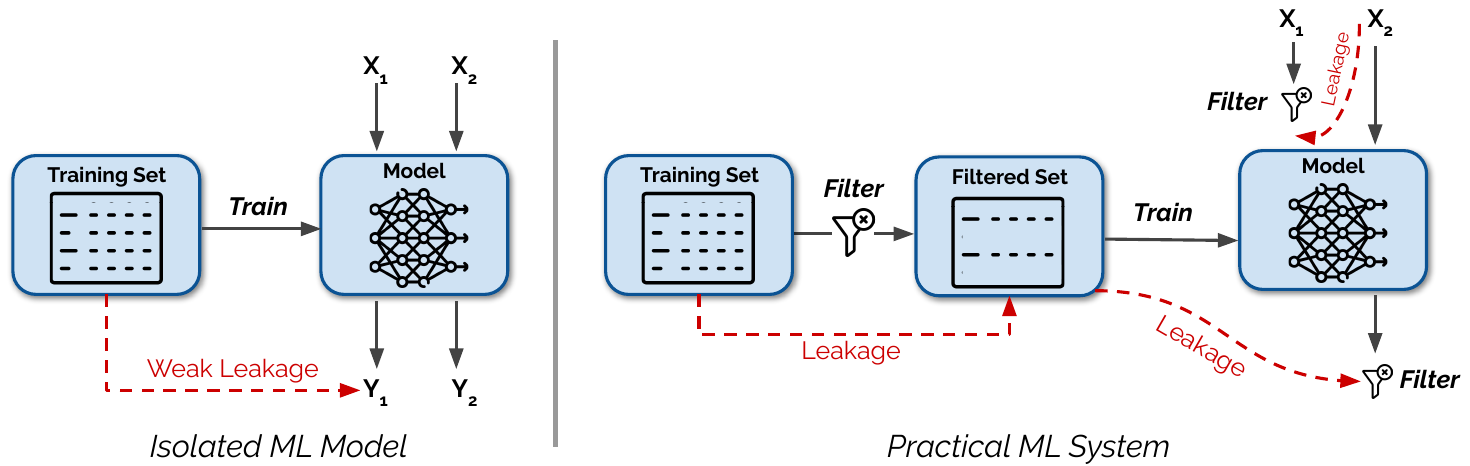}
    \vspace{-0.1cm}
    \caption{Most past work studies the privacy leakage of ML models in \textit{isolation}. However, in reality most models are part of larger \textit{systems} that contain 
    components for filtering training data, blocking certain test inputs, monitoring model outputs, and more. In this work, we show that adaptive adversaries can create side channels that exploit these system-level components to significantly amplify privacy leakage.}
    \label{fig:main}
    \vspace{-0.3cm}
\end{figure*}

\item \textbf{Input preprocessing (Section~\ref{sec:test_filters}).} Many models require their inputs to be preprocessed, e.g., language models require text to be tokenized. When these preprocessors are built using training statistics (e.g., tokenizers), we show that it creates side channels that allows adversaries to extract private information such as rare training words~\citep{rumbelow23solidgoldmagikarp}.

\item \textbf{Model output filtering (Section~\ref{sec:test_filters}).} To improve privacy,
many ML systems include filters that prevent the system from outputting verbatim training data~\citep{githubcopilot,ippolito2022preventing}. We show that this creates a side channel that actually \textit{reduces} privacy, even enabling near-perfect membership inference. 
We use this attack to reverse engineer details of GitHub Copilot's training set (namely, its exact training data cutoff)
and then use it to extract OpenSSH private keys from a public language model's training set.

\item \textbf{Query filtering (Section~\ref{sec:query_filters}).} Many ML systems use test-time query filters that reject certain inputs, e.g., detectors for adversarial examples~\citep{li2020blacklight, rashid2023malprotect, maghsoudimehrabani2022proactive} and model extraction attacks~\citep{juuti2019prada, zhang2021seat, pal2020activethief, liu2022seinspect}. We show that since many of these filters aggregate information across different users (to protect against Sybil attacks), adversaries can reveal information about other users' queries by sending targeted inputs. Unlike the previous side channels that reveal private \textit{training} examples, this approach reveals other users' private \emph{test} queries, which is otherwise impossible when considering isolated ML models.
\end{itemize}

Taken together, our results highlight the need to shift to a system-level view when analyzing ML privacy. Unfortunately, we show that guaranteeing system-level privacy is challenging, both empirically and theoretically, due to highly non-trivial interactions among different components. For example, we show that combining DP and deduplication (as suggested by past work~\citep{ponomareva2022training}) leads to worse privacy than DP on its own. Furthermore, system-level analyses require tackling the trade-offs that arise when we consider a model's trustworthiness across \emph{multiple} axes, e.g., query filters that improve security can simultaneously hurt privacy.
Overall, we hope that our work can contribute to future improvements on the frontiers of ML privacy.

\section{Preliminaries}
\label{sec:prelims}

\subsection{Background and Related Work}
\label{ssec:background}

Our paper studies a wide range of attacks that exploit numerous ML models and system components. In this section, we discuss general preliminaries for ML systems and privacy attacks; we otherwise defer the relevant background to each self-contained attack section (Sections \ref{sec:train_filters}--\ref{sec:query_filters}).

\subsubsection{Standalone ML Models}
\label{ssec:standalone}
An ML model is a function $y \gets f(x)$ that is trained on a dataset $D_{\mathrm{train}}$ and evaluated at inference time on a sequence of queries $D_{\mathrm{test}} = \{x_1, x_2, \dots \}$.
Most past work on ML privacy considers a ``standalone'' or ``isolated'' ML model where $f$ is trained directly on $D_{\mathrm{train}}$ (e.g., using gradient descent) and then independently queried on each data point $x_i \in D_{\mathrm{test}}$ to get outputs $y_i \gets f(x_i)$.

\paragraph{Privacy attacks and defenses}
When considered as standalone functions, ML models can memorize and leak considerable information about their training data. Numerous attacks exploit this leakage, including \emph{membership inference attacks} \citep{shokri2017membership, yeom2018privacy,choquette2021label, carlini2022membership} that infer whether a specific example was in the training set, and more powerful \emph{data extraction attacks} \citep{carlini2019secret, carlini2021extracting, balle2022reconstructing, carlini2023extracting, somepalli2023diffusion, nasr2023scalable} that can reveal entire training examples.
To defend against these attacks, models can be trained with \emph{differential privacy}~\citep{dwork2006calibrating}---typically
by clipping and adding noise to updates in stochastic gradient descent~\citep{abadi2016deep,choquette2022multi}---which makes them provably secure against privacy leakage.
Other more heuristic approaches such as \emph{data deduplication}~\citep{lee2021deduplicating, kandpal2022deduplicating} or \emph{output filtering}~\citep{githubcopilot} can also mitigate some (but not all) memorization~\citep{ippolito2022preventing, carlini2022quantifying, kandpal2022deduplicating}.

\paragraph{Membership and non-membership inference} Many of our side-channels exploit membership inference attacks, where the goal is to infer whether a specific data sample $x$ was used in the training of a model $f$.
In real-world scenarios, it is important for attackers to have low false-positive rates as most arbitrary samples are non-members~\citep{carlini2022membership}. 
In addition, inferring \emph{non}-membership is significantly easier than inferring membership, since ML models typically make no mistakes on their training data (i.e., if a model gets an example wrong then it is probably a non-member).
For many of our attacks, we show that side channels can be used to convert a membership inference problem into an equivalent (and easier) non-membership inference problem.%

\subsubsection{ML Systems}
\added{We consider part of an ML \emph{system} everything that composes the data collection, training and  deployment pipeline. These components} augment the standalone model in \Cref{ssec:standalone} with additional components that act upon model inputs or outputs. \added{In particular}, we consider the following generic system components (see \Cref{fig:main} for an illustration):

\begin{itemize}[topsep=2pt, itemsep=0pt, leftmargin=10pt]
    \item \textbf{A training data filter} (Section~\ref{sec:train_filters}) is a function  %
    that modifies or removes inputs from the data before training.
    Popular examples include outlier removal or deduplication.
    \item \textbf{An input pre-processor} (Section~\ref{sec:test_filters}) is a function that modifies inputs to prepare them for running through the model. Popular examples include image cropping, or tokenization and truncation for language models.
    \item \textbf{An output post-processor} (Section~\ref{sec:test_filters}) is a function %
    that acts on the output $y \gets f(x)$ produced by the trained model and returns a modified output $y'$ (possibly an empty output $y' = \bot$).
    Most relevant to our work are memorization filters that reject outputs $y$ that leak information about some training example.
    \item \textbf{A query filter} (Section~\ref{sec:query_filters}) is a function %
    that acts on the sequence of input queries $D_{\mathrm{test}}$ sent to a trained model and returns modified inputs $D_{\mathrm{test}}' = \{x_1', x_2', \dots \}$. The function may reject some input queries (i.e., $x_i' = \bot$) in which case the system returns no output for that query.
    Common examples include methods for rejecting inputs that are perceived as being part of some attack attempt.

\end{itemize}

\paragraph{\added{Privacy} side channels in ML systems}
We broadly define a privacy \emph{side channel} in an ML system as any instance where an attacker can probe the system-level components to leak significantly more information about the training set $D_{\mathrm{train}}$ or query set $D_{\mathrm{test}}$ than what a similarly capable attacker could leak by interacting with a standalone model.
This additional leakage is made possible when instantiations of the above filters introduce \emph{dependencies} between different model inputs---either at training or inference time.

We do not consider modifications to the training algorithm \citep{song2019privacy} or to the training data \citep{tramer2022truthserum} to be ``side channels'' as they also apply to the isolated model case. Similarly, other types of attacks such as exploiting ML models used for computer systems~\citep{schuster2022learned} (e.g., caches, databases, etc.) are out-of-scope for our study.

\added{We note that prior work has considered the impact of system components on the \emph{integrity} of ML model deployments, e.g., by abusing image preprocessing algorithms to create adversarial examples~\citep{xiao2019seeing,quiring2020adversarial}.}

\subsubsection{A Comparison to Physical Side Channels}
In cryptography and computer security, side-channel attacks are typically defined as attacks that exploit leakage from the \emph{implementation} of a protocol or algorithm, rather than from the design itself.
Side-channel attacks often rely on leakage through physical phenomena (e.g., power consumption or timing~\citep{kocher1996timing, kocher1999differential})
or auxiliary functionalities such as error messages~\citep{bleichenbacher1998chosen, vaudenay2002security}.
In the context of machine learning, physical side-channel attacks can recover the weights or architecture of standalone ML models, e.g., via electromagnetic signals~\citep{batina2019csi}, timing~\citep{duddu2018stealing}, power consumption~\citep{xiang2020open}, GPU context-switches~\citep{wei2020leaky}, memory access patterns~\citep{hu2019neural}, \added{fault injections~\citep{breier2022sniff}, or cache timing~\citep{hong2018security,yan2020cache}}. In addition, one can reconstruct model queries at inference time using power analysis~\citep{wei2018know}.

Our work considers a different form of side channel that does not rely on physical signals, but instead relies on leakage introduced by auxiliary functionalities present in real-world deployments of machine learning. This is analogous to how deploying secure encryption alongside functionalities like error reporting enables new side channel attacks~\citep{bleichenbacher1998chosen, vaudenay2002security}), or how deduplication can harm cloud storage systems~\citep{halevi2011proofs}.
In contrast to physical side-channel attacks, our attacks do not require any physical access or fine-grained measurement of the ML model. Instead, we only require black-box access to the system's prediction function.

\subsection{Threat Model}
\label{ssec:threat_model}

\paragraph{Adversary's goal} The adversary's goal is to leak information about the training data $D_{\mathrm{train}}$ or test queries $D_{\mathrm{test}}$
of an ML system by leveraging system-level side channels.
For membership inference attacks, we consider the standard adversarial game where we pick a data point $x$, include this sample in $D_{\mathrm{train}}$ with 50\% probability, and then have the attacker predict whether $x$ was in $D_{\mathrm{train}}$.
We also consider test-time membership inference attacks where the attacker has to guess whether some query $x$ has previously been made to the system or not, i.e., if $x\in D_{\mathrm{test}}$.

\paragraph{Adversary's system access} We assume that the attacker has only black-box access to the ML system, i.e., they can send arbitrary queries and observe outputs, but they do not know the exact weights of the ML system. On the other hand, we assume the attacker knows which system components %
are used (e.g., output filters) and how they are implemented (for some of our attacks, this knowledge is not strictly necessary). We believe this assumption is reasonable for three reasons:
(1) we want to distinguish privacy that may be inherent in a system, from ``privacy-through-obscurity'' that arises solely due to (likely short-lived) secrecy of some system configurations;
(2) many system components we consider have a small number of standardized implementations that practitioners are likely to use. \added{For example,} \texttt{imagededup}\footnote{\url{https://github.com/idealo/imagededup}} and \texttt{DataTrove}\footnote{\url{https://github.com/huggingface/datatrove}} \added{are the most widely used implementations for data deduplication of images and text, respectively, and offer only one or two canonical methods;}
(3) some of our attacks are reasonably robust to uncertainty and work even without precise knowledge of the system: for instance, our attack against GitHub Copilot works even though the system is a black-box.

In some of our attacks (Section~\ref{sec:train_filters}, \ref{sec:dp}), we assume that the attacker can poison the training set, roughly mirroring the setting of \citet{tramer2022truthserum}. In our data extraction attack against language models (Section~\ref{ssec:vocab_extraction}), the attacker is assumed to know the prefix $p$ of a string $p||s$ from the training set.
For all other attacks, we make no additional assumptions on the adversary.

Since we are introducing a new class of attacks, we generally assume that there are no explicit defenses in place against privacy side channels. However, many of our attacks operate in settings where a model developer has deployed provable or heuristic defenses against privacy leakage (e.g., differential privacy, data deduplication, or memorization filters).

\subsection{Ethics and Broader Impact}
The attacks that we present could present a threat to user privacy in deployed ML systems. To mitigate these harms, we either study systems that we create ourselves using public data (e.g., CIFAR-10 classifiers) or ones such as GitHub Copilot that are trained on publicly-available data and thus do not pose any real-world privacy risks. The goal of our work is to bring to light these privacy vulnerabilities in order to spur future work on developing private end-to-end systems.

\section{Attacking Training Data Filters}
\label{sec:train_filters}

We first study side-channel attacks on training data filters. 
We show that common data filters introduce strong \emph{co-dependencies} between data samples: a data point might be filtered out if and only if some other data points are present in the training set.
This introduces a side channel where an adversary can determine, with high confidence, whether a target data sample was present in the training set. %

\added{To isolate the privacy leakage that stems from the addition of a specific filter, we assume throughout this section that the system applies a single data filtering scheme before training.}

\subsection{Attacking Training Data Deduplication}
\label{ssec:dedup}

Most large-scale ML models are trained on data scraped from the Internet~\citep{schuhmann2022laion, gao2020pile}.
Due to the nature of Internet content, it is common for some training data instances to be repeated multiple times;
this is problematic for several reasons, including that
repeated training data is much more likely to be memorized~\citep{carlini2022quantifying}.
The process of \emph{deduplication} addresses this challenge by removing exact or near-duplicated samples~\citep{lee2021deduplicating}.
Most modern large language models and image generation models use training data deduplication, e.g., Gopher~\citep{rae2021scaling}, DALLE-2~\citep{ramesh2022hierarchical}, LLaMA 2~\citep{touvron2023llama}, PaLM 2~\citep{anil2023palm}, and others.

\paragraph{Types of data deduplication}
We primarily focus on deduplication applied to images, although the principles underlying our attacks are directly applicable to other domains such as text, as we show in \Cref{ssec:dedup_variants}.
As in \citet{lee2021deduplicating}, we consider both \emph{exact} and \emph{approximate} deduplication.
Exact duplicates are samples $x, x'$ that are equivalent, i.e., $x=x'$. Approximate duplicates are samples $x, x'$ that are ``close'', i.e., $\texttt{sim}(x, x') \geq \alpha$ for some similarity measure $\texttt{sim}$ and threshold $\alpha$.
Once duplicates are identified, there are two approaches for deduplication: \emph{delete-all} and \emph{delete-all-but-one}. If $x\in D_{\mathrm{train}}$ has as least one duplicate $x'\in D_{\mathrm{train}}$, then \emph{delete-all} removes all duplicates $x,x'$. \emph{Delete-all-but-one} keeps exactly one copy, i.e., only one of $x'$ and $x$ is retained. While the latter may seem more natural, the former is simpler to implement; both approaches have been proposed in prior work and used in practice~\cite{lee2021deduplicating}.

\paragraph{The side channel intuition}

Data deduplication introduces co-dependencies between data points: a sample $x$ is deleted if and only if some similar sample $x'$ is present in the training set.
An attacker can exploit this side channel to perform a strong targeted membership inference attack.
Concretely, suppose the attacker wants to infer membership of some sample $x$ and knows that some near-duplicates $x'$ are in $D_\textrm{train}$. Then, the adversary can infer if $x$ is in $D_\textrm{train}$ by detecting whether the duplicates $x'$ were deleted or not.
Compared to a standard membership inference attack on the targeted point $x$, this side channel amplifies the privacy leakage if the duplicates $x'$ known to the adversary are more likely to be memorized than the original data point.

\subsubsection{Our Attack}
We operate in the same threat model as \citet{tramer2022truthserum}: the attacker can \emph{poison} a small fraction of the dataset to leak information about other data points.
In our case, the adversary poisons the training set by inserting \emph{duplicates} $x'$ of a targeted data point $x$. 
We assume the attacker has black-box query access to the trained model $f$ and they know the data deduplication procedure that is used. 

We focus on classification tasks with labeled data $(x,y)$ and extend our attack to language modeling in \Cref{ssec:dedup_variants}.
We assume that deduplication is independent of a sample's label, i.e., two points $(x,y)$ and $(x', y')$ are duplicates if $x$ is close to $x'$, even if $y\neq y'$.
As in \citet{tramer2022truthserum}, we assume the attacker can introduce mislabeled data points into the training set. This assumption could be relaxed in some settings by considering clean-label poisoning attacks~\citep{shafahi2018poison}. \change{We further assume that the model is trained on all (deduplicated) data (i.e., there is no data subsampling step).}

\begin{figure}[t]
    \centering
    \begin{minipage}{0.2\textwidth}
    \includegraphics[scale=.5]{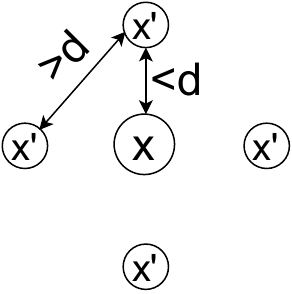}
    \end{minipage}
    \begin{minipage}{0.26\textwidth}
    \includegraphics[width=\textwidth]{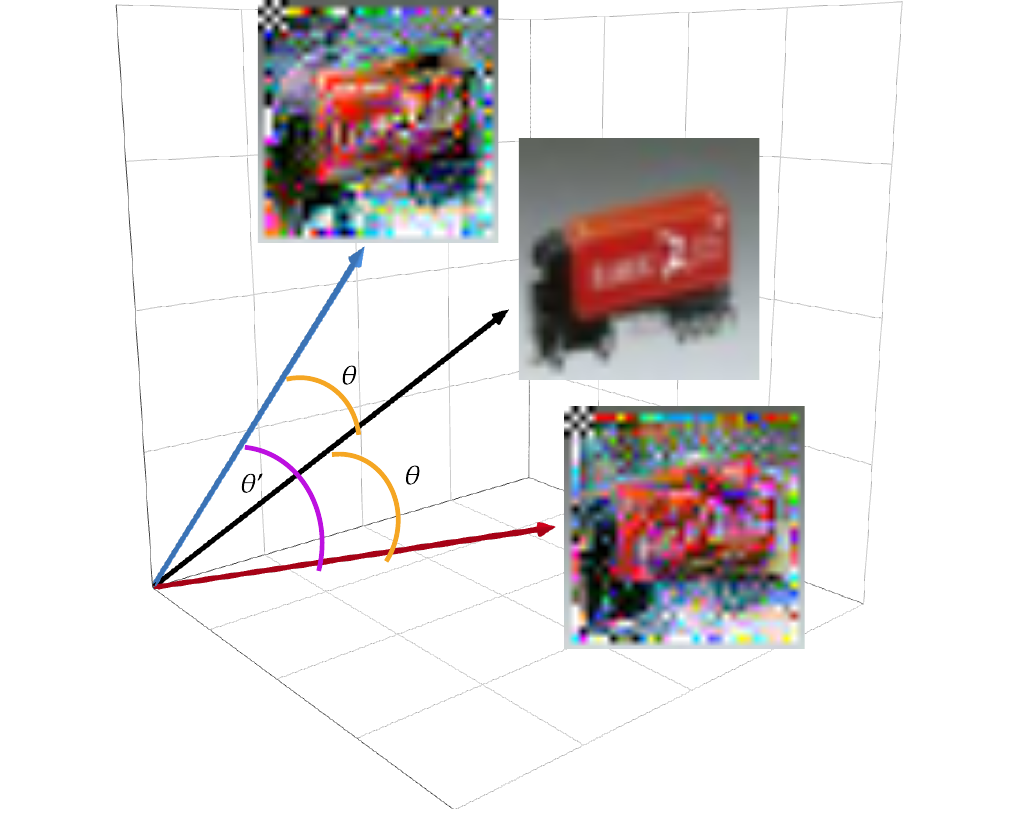}
    \end{minipage}
    \caption{A depiction of our ``hub-and-spokes'' attack on data deduplication. \textbf{Left:} we insert poisoned examples that are each close to the ``hub'' ($x$) but are far from each other.
    \textbf{Right:} actual images from our attack. We also include a checkerboard backdoor in the top-left corner of each near-duplicate image to enhance memorization.}
    \label{fig:approx-dup}
    \vspace{-0.5em}
\end{figure}

We propose attacks for the four forms of deduplication that we introduced above:

\begin{itemize}[topsep=2pt, itemsep=0pt, leftmargin=15pt]
    \item \emph{Exact deduplication, delete all}: Given a target $(x, y)$, we add a mislabeled duplicate $(x, y')$ into the dataset. We then run \emph{non-membership} inference on the mislabeled point: if this point is absent, we know that the target was present before deduplication. This attack only requires inserting a single poison example.
    
    \item \emph{Exact deduplication, delete all-but-one}: In this setting we still use the same attack as above, but the attack is less powerful: when the target is a member, the poisoned duplicate only gets removed with 50\% chance. This attack also only requires inserting a single poison example.
    
    \item \emph{Approximate deduplication, delete all-but-one}: This is the most complex and interesting attack setting. We create $N$ approximate duplicates $(x'_1,y'), \dots, (x'_N,y')$ in a ``hub-and-spoke'' pattern (described further below and illustrated in \Cref{fig:approx-dup}). This causes the attacker's points to be near-duplicates of the target, i.e., $\texttt{sim}(x, x'_i) \geq \alpha$ for all $i$, but not of each other, i.e., $\texttt{sim}(x'_i, x'_j) < \alpha$ for all $i \neq j$. The attacker then runs a non-membership inference attack across \emph{all} $N$ poisoned samples $x'_i$.
    In this way, the attack becomes stronger as the attacker introduces more poison examples.
    \item \emph{Approximate deduplication, delete all}: We repeat the same approximate deduplication attack, but the attack is stronger as all examples are removed.%
\end{itemize}

We observe that the side-channel cannot be trivially removed via dataset pre-processing. Manual filtering is too impractical for modern, large datasets, and automated filtering would not be able to remove the side-channel: a filter removing equal samples with differing labels would lead to a scenario equivalent to the \textit{delete all-but-one} deduplication described above. \added{Also randomly subsampling the training data would not be enough to defend from our attack against approximate deduplication with multiple poisons (even though it might be slightly weaker). However, the attack with a single duplicate may fail.}

\paragraph{Generating approximate duplicates} 
In our experiments, we use a standard approximate deduplication procedure based on the cosine similarity between the embeddings $h(x), h(x')$ of two samples, where $h$ is a neural network. Assuming that $h$ outputs embeddings of unit norm in $\mathbb{R}^d$, we have $\texttt{sim}(x, x') \coloneqq h(x) \cdot h(x')$.
We thus want to find duplicates $x'_1, \dots, x'_N$ of a target $x$ such that $h(x) \cdot h(x'_i) \geq \alpha$ for all $i$, and $h(x'_i) \cdot h(x'_j) < \alpha$ for all $i \neq j$.
To find approximate duplicates, we first compute embeddings that satisfy the above condition, and then ``invert'' the embedding function $h$ to find input images.

Assume without loss of generality that the target embedding $h(x)$ is the first basis vector in $\mathbb{R}^d$, i.e., $h(x) \coloneqq (1, 0, \dots, 0)$ (for a general unit-norm embedding $h(x)$, we simply have to apply an appropriate rotation to all vectors). We now build $d-1$ unit vectors $e_1, \dots, e_{d-1}$ of the form
\setlength{\abovedisplayskip}{3pt}
\setlength{\belowdisplayskip}{4pt}
\[
e_i \coloneqq (\alpha, \underbrace{0, \dots, 0}_{i-1}, \sqrt{1 - \alpha^2}, 0, \dots, 0) \;.
\]

This ensures that: (1) all the embeddings $e_i$ are of unit norm; (2) the cosine similarity between each embedding $e_i$ and the target $h(x)$ is $\alpha$; (3) the cosine similarity between each pair of near-duplicate embeddings is $\alpha^2 < \alpha$.
Given these embeddings $e_1, \dots, e_N$, we invert the embedding function $h$ by running 1000 steps of projected gradient descent. That is, we optimize the input image $x'_i$ so as to maximize the cosine similarity between $h(x'_i)$ and $e_i$.

To further maximize the membership inference signal, we make the mislabeled near-duplicates $(x'_i, y')$ to be as easy to memorize as possible.
We achieve this by adding a common backdoor feature to all near-duplicates (a checkerboard pattern as in~\citet{gu2017badnets}).
In the extended version of this paper \cite[Appendix A]{debenedetti2023privacy} we show that including this backdoor pattern significantly strengthens our attack. In \cref{fig:approx-dup}, we present an example depiction of our approximate duplicate attack.

\paragraph{Non-membership inference attack} After poisoning, we adapt the LiRA method~\citep{carlini2022membership} to run membership inference. In particular, we train ``shadow models'' that either contain, or do not contain, the attacker's poisoned duplicates. We query all the shadow models to obtain the models' confidences on the duplicates' poisoned label, and then fit two multivariate Gaussians (one when the duplicates are members and one when they are not) on the model's confidence on each duplicate.

For the final attack, we query the target model $f$ on each duplicate to obtain the confidence in the poisoned label, and then perform a Gaussian likelihood ratio test to determine whether the duplicates are likely to be members of the deduplicated data.
If the duplicates are predicted as non-members of the deduplicated data, then we predict that the target example $x$ was in the original training set.%

\subsubsection{Evaluation}
We evaluate our attacks on standard models trained on CIFAR-10 and deduplicate the images using the popular open-source \texttt{imagededup} Python library. We use the library's default settings, which compares images by the cosine similarity of their embeddings computed from a pretrained MobileNetV3 model~\citep{howard2019searching}. Two images are considered duplicates if the similarity is at least $\alpha=0.9$.

We choose 250 targets at random from the CIFAR-10 training set and create between 1 and 8 exact- or near-duplicates per target. We build the training set $D_{\mathrm{train}}$ by randomly sampling $50\%$ of the CIFAR-10 training set and adding the duplicates. We train the models for 100 epochs.

Following \citet{carlini2022membership}, we evaluate the success rate of the membership inference attack by its true-positive rate (TPR) at a low false-positive rate (FPR).
As baselines, we consider their original LiRA attack as well as the attack of \citet{tramer2022truthserum} that combines LiRA with data poisoning (similar to our attack, but not leveraging the side-channel).

\begin{figure}[t]
    \centering
    \includegraphics[width=0.4\textwidth]{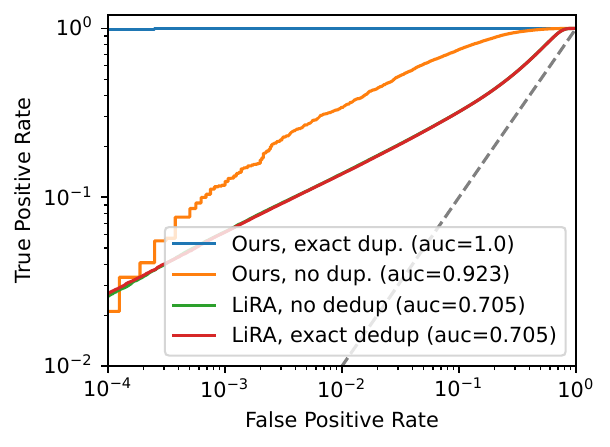}
    \caption{\textbf{Deduplication can significantly worsen privacy.} 
    We show membership inference effectiveness under both exact deduplication (\emph{delete all}) and no deduplication. With deduplication, the side-channel leads to near-perfect membership inference; without, it is similar to the baseline poisoning-aware Truth Serum attack~\citep{tramer2022truthserum}.
    The LiRA baseline~\citep{carlini2022membership} performs similarly in both cases.}
    \label{fig:dedup_vs_no_dedup}
    \vspace{-1.0em}
\end{figure}

\subsubsection{Results}
Deduplication creates a strong side channel that enables near-perfect membership inference.
In \cref{fig:dedup_vs_no_dedup}, we compare membership inference attacks against models trained with and without delete-all, exact data deduplication.
On deduplicated data, our attack achieves essentially perfect targeted membership  inference, with a TPR of $98\%$ at a FPR of $0.01\%$.
When data deduplication is not applied, our attack still outperforms LiRA because of the amplification that poisoning has on membership inference, as shown by \citet{tramer2022truthserum}.

In \cref{tab:dedup-del-all}, we provide a more comprehensive comparison between our attack and the prior state-of-the-art attack (Truth Serum~\citep{tramer2022truthserum}) under different deduplication settings.
Although both attacks leverage poisoning by introducing mislabeled samples, only ours explicitly exploits the deduplication side-channel.
For the single poison case, both our attack and Truth Serum add a mislabeled copy of the target to the training set.
However, our attack differs in how we perform inference: we perform \emph{non-membership inference} on the \emph{mislabeled} duplicate, while Truth Serum performs membership inference on the correctly labeled target sample. 
The latter is sub-optimal when deduplication is applied, since the correctly labeled target is \emph{never} in the training set. In the \emph{delete all} deduplication setting, this change boosts the attack's TPR to near 100\%.

In the \emph{delete all-but-one} setting, our attack has lower overall TPR (but still outperforms Truth Serum) due to the possibility that one of the poisons survives deduplication even when the target is a member.
With approximate deduplication, our hub-and-spoke attack design with 8 duplicates doubles the attack's TPR compared to using a single poison.
Note that in contrast to our attack, running Truth Serum with $>\!\!1$ poisons completely breaks the attack when deduplication is applied. This is because Truth Serum's poisons are all exact (mislabeled) copies of the target, and thus, deduplication removes the poisons regardless of whether the target is in the training set or not.
Our attack sidesteps this issue by ensuring that the attacker's poisons are near-duplicates of the target, but \emph{not} near-duplicates of each other.

\begin{table}[t]
\centering
\caption{\textbf{Our attacks improve prior privacy-poisoning attacks.} We report the TPR at a FPR of 0.1\% and 0.01\%, and we compare our attacks against Truth Serum~\citep{tramer2022truthserum} and LiRA~\citep{carlini2022membership} under different deduplication strategies. In parentheses, we report the number of poisons injected.}
\vspace{0.5em}
\label{tab:dedup-del-all}
\renewcommand{\arraystretch}{0.5}
\resizebox{\columnwidth}{!}{\begin{tabular}{@{}llrr@{}}
\toprule
&& \multicolumn{2}{c}{TPR}\\
\cmidrule(l{5pt}){3-4}
Deduplication & Attack                  & @ 0.1\% FPR & @ 0.01\% FPR \\ \midrule
\multirow{3}{*}{None} & LiRA                    & $6.2\%$                            & $2.6\%$                             \\
                                  & Truth Serum (1)  & $18.0\%$                           & $4.0\%$                             \\
                                  & Truth Serum (8) & $\mathbf{30.5\%}$                            & $\mathbf{6.3\%}$                              \\
\midrule
\multirow{4}{*}{\shortstack[l]{Exact or Approx\\\emph{delete all}}}            & LiRA                    & $6.2\%$                            & $2.7\%$                             \\
                                  & Truth Serum (1)  & $89.8\%$                           & $79.3\%$                            \\
                                  & Truth Serum (\textgreater 1) & $0.1\%$                               & $0.0\%$                                \\
                                  & Ours (1)         & $\mathbf{99.6\%}$                           & $\mathbf{98.0\%}$                            \\ \midrule
\multirow{4}{*}{\shortstack[l]{Exact\\\emph{delete all-but-one}}}                                   
& LiRA & $6.2\%$ & $2.7\%$ \\
& Truth Serum (1) & $41.6\%$ & $31.8\%$ \\
& Truth Serum (\textgreater 1) & $3.1\%$ & $1.1\%$ \\
& Ours (1) & $\mathbf{45.6\%}$ & $\mathbf{40.9\%}$ \\ 
\midrule
\multirow{3}{*}{\shortstack[l]{Approximate\\\emph{delete all-but-one}}}      & LiRA                    & $6.5\%$                            & $2.9\%$                             \\
                                  & Truth Serum (\textgreater 1) & $0.1\%$                               & $0.0\%$                                \\
                                  & Ours (8)        & $\mathbf{96.0}\%$                           & $\mathbf{90.4\%}$                             \\ \bottomrule
\end{tabular}}
\end{table}

\subsubsection{\added{Dealing with Attacker Uncertainty}}

\added{In practice, an attacker might not know exactly which deduplication method is being used, or with which hyperparameters (for approximate deduplication). Here, we show that an attacker can still implement a successful attack in such a case.}

\paragraph{Unknown deduplication method} \added{In this case, the attacker should apply our attack that adds one mislabeled exact duplicate. Note that this attack is assumption-free: it requires no knowledge of the deduplication approach that is used. This strategy is optimal for three-out-of-four deduplication settings: Exact \emph{delete all}, Approx \emph{delete all}, Exact \emph{delete all-but-one}. The one setting where this attack is not optimal is for the Approx \emph{delete all-but-one} setting. However, in this case the attack with one mislabeled exact duplicate still outperforms Truth Serum, the prior state-of-the-art (the results are the same as for the Exact \emph{delete all-but-one} setting, where our attack has a $9\%$ higher TPR @ $0.01\%$ FPR compared to Truth Serum, as shown in \Cref{tab:dedup-del-all}).}

\paragraph{The attacker knows Approx \emph{delete all-but-one} is used} \added{If the adversary knows that the system is using Approx \emph{delete all-but-one} deduplication, 
and they can guess the $\alpha$ reasonably well, then they can improve over the assumption-free attack above. If the attacker’s guess for $\alpha$ is close to correct, the attack is near-perfect ($96.0\%$ TPR @ $0.1\%$ FPR) by adding eight approximate duplicates.
Specifically, the attacker’s guess must satisfy $\alpha_{guess}^2 < \alpha_{real} \le \alpha_{guess}$. If the attacker’s guess does not satisfy the above, or if the deduplication method is not actually Approx \emph{delete all-but-one}, the attack fails.
Thus, if the attacker is not confident they can guess $\alpha$ correctly, or they are uncertain if Approx \emph{delete all-but-one} is used, they should just use the assumption-free attack above, and still outperform Truth Serum.}

\paragraph{How can the adversary guess \texorpdfstring{$\alpha$?}{alpha}}

\begin{figure}[t]
    \centering
    \includegraphics[width=0.4\textwidth]{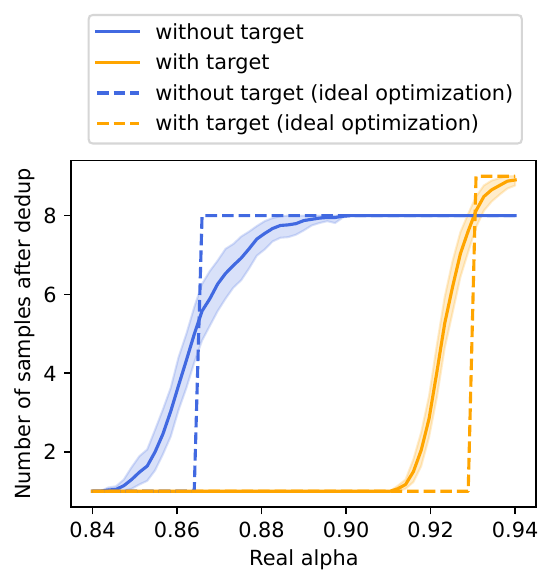}
    \caption{\added{\textbf{Our attack is robust to uncertainty in the deduplication threshold $\alpha$.} We create approximate duplicates by optimizing for the poisons to be less than $\alpha^2$-similar to each other, and more than $\alpha$-similar to the target. We show that, within a $\pm 2\%$ range, six duplicates are kept on average when the target sample is not a member--with only one sample being kept when the target sample is a member.}}\label{fig:alpha-range}
    \vspace{-1.0em}
\end{figure}

\added{The adversary can estimate $\alpha$ by computing a TPR-FPR curve over a sample of training data. We estimate that $\alpha$ is unlikely to be smaller than $0.8$ in practice as this would yield too many false positives, and reduce the size of the training set. Moreover, the attacker does not need to guess $\alpha$ \emph{exactly}. Even if the guess is off by $\pm 2\%$, most of the poisons will survive the deduplication process, as shown in \Cref{fig:alpha-range}. Crucially, this rules out $\alpha$ randomization as a potential defense against our attack. Finally, if the same deduplication method is reused multiple times, the attacker can recover the real alpha value by adding dummy near-duplicates with different similarities. Thus, if a developer collects multiple training sets over time and keeps the same deduplication process, the attacker can learn the parameters once and then use this to reliably attack future training runs.}

\subsubsection{Extension to the Text Domain}
\label{ssec:dedup_variants} 
While we have focused on classification tasks so far, our attack techniques can easily be extended to other tasks such as language modeling.
Text data is commonly deduplicated by finding and removing substrings of $k$ tokens that are exactly repeated across documents~\citep{lee2021deduplicating}.
We show this method creates a side channel that enables a powerful \emph{attribute inference attack} (a more powerful attack than membership inference).
Specifically, we show that the attacker can infer a low-arity attribute (e.g., a medical condition) contained within a known piece of training data (e.g., a template of a healthcare form). For simplicity of exposition, we assume the unknown attribute is a single token, but our attack could be generalized to longer attributes.
We assume the training set contains a sentence of the form
\[
\underbrace{{\color{brown}S_1}, \dots, {\color{brown}S_{k-1}}}_{\color{brown}\texttt{pre}}\  {\color{red}\underline{X}}\  \underbrace{{\color{brown}S_{k+1}}, \dots, {\color{brown}S_{2k-1}}}_{\color{brown}\texttt{post}} \;,
\]
where $\texttt{pre}$ and $\texttt{post}$ are a prefix and suffix of lengths $k-1$ known to the attacker, and ${\color{red}\underline{X}}$ is an unknown token that comes from a set 
of $N$ possible values $\{\attribute_1, \dots, \attribute_N\}$. The attacker's goal is to infer the value of ${\color{red}\underline{X}}$.

\paragraph{Our attack}
At a high level, our attack creates $N$ families of poison strings (with $k$ poison strings per family), and where each family's strings contains some ``canary'' substring that is deduplicated if and only if the true value of the unknown attribute is $\attribute_i$.
Then, by running an inference attack to detect the presence of these canary substrings, the attacker can infer which attribute value is correct.

More concretely, we build the i-th family by choosing some unique strings ${\color{blue} A_i, B_i}$ (each of length less than $k$) and building $k$ strings of the form:
\begin{alignat*}{5}
    {\color{blue}A_i}\ +\ &{\color{brown}S_1} &&{\color{brown}S_2}&&{\color{brown}S_3} \dots {\color{brown}S_{k-1}} &&\ {\color{red}\attribute_i}\  &&\ +\ {\color{blue}B_i}\\
    {\color{blue}A_i}\ +\ & && {\color{brown}S_{2}} &&{\color{brown}S_3} \dots {\color{brown}S_{k-1}} &&\ {\color{red}\attribute_i}\  {\color{brown}S_{k+1}} &&\ +\  {\color{blue}B_i}\\
    {\color{blue}A_i}\ +\ & && &&{\color{brown}S_3} \dots {\color{brown}S_{k-1}} &&\ {\color{red}\attribute_i}\  {\color{brown}S_{k+1}} {\color{brown}S_{k+2}} &&\ +\  {\color{blue}B_i}\\
    & && && && \ddots\\
    {\color{blue}A_i}\ +\ & && && &&\ {\color{red}\attribute_i}\  {\color{brown}S_{k+1}} \dots {\color{brown}S_{2k-1}} &&\ +\  {\color{blue}B_i} \;.
\end{alignat*}

Our construction ensures that: (1) none of the adversary's $N \cdot k$ poison strings contain a duplicated substring of length $k$; (2) if the $i$-th family of poison strings has the correct value of the unknown attribute ${\color{red}\underline{X}}$, all strings in the family are deduplicated to the string ${\color{blue} A_i+ B_i}$ (we assume ${\color{blue} A_i+ B_i}$ to be shorter than $k$ so that it is not deduplicated recursively).

Once the language model has been trained on deduplicated data, the adversary can compute the model's loss on each of the $N$ strings ${\color{blue} A_i+ B_i}$. The string with the lowest loss likely corresponds to the index of the unknown attribute, as these strings only appear contiguously in the training set if deduplication has occurred.

\subsubsection{Deduplication Is Still Worth It in Practice} Deduplication is a cornerstone in mitigating memorization in deployed ML models
trained on web-scale datasets~\citep{kandpal2022deduplicating,lee2021deduplicating}.
Moreover, differentially private training is only effective on deduplicated data (or else privacy
leakage scales exponentially with the number of duplicates).
However, if deduplication can induce \emph{new} privacy violations as we show above, should practitioners still apply it?
We argue yes, because the empirical cost of \emph{not} doing so is just too high---the privacy risks of duplicated data are far too egregious.
This highlights an unfortunate trade-off, where the best practice of using deduplication will be vulnerable to adversarial attacks, and we thus hope that new techniques can be developed to deduplicate datasets without aggravating
privacy leakages.

\subsection{Attacking Poisoning Defenses}
\label{ssec:poisoning_defenses}
We next look at a broader class of data filtering techniques that aim to protect against poisoning attacks~\citep{biggio2012poisoning}. 
Poisoning attacks introduce malicious behavior into a model, often by adding many samples that share a common feature (e.g., a backdoor~\citep{chen2017targeted}).
Defenses against poisoning often work by removing examples based on their relationships to other examples and thus they create ``codependencies'' that can be targeted to leak information.

\subsubsection{Poisoning Defenses}
Defenses against poisoning attacks find and remove clusters of points that are part of a presumed attack.
Many defenses end up performing some form of data deduplication and are thus vulnerable to a similar side channel as presented in the previous section. 

As a straightforward example, \citet{carlini2021poisoning} propose a defense against data poisoning that detects poisoned samples and also occasionally flags near-duplicates (as false-positives). Thus, deploying this defense would immediately create the exact same privacy side-channel as described in the previous section.

Alternatively, Sybil filtering~\citep{fung2018mitigating,awan2021contra,nguyen2022flame} is a common countermeasure to prevent colluding adversaries from poisoning a federated learning (FL) model \citep{mcmahan2017communication}.
Most implementations of Sybil filtering downscale or remove a client's model updates if they are too similar to another client's updates. The defense's intuition is that colluding clients performing a poisoning attack are likely to contribute very similar updates, while benign clients' updates are likely to be diverse due to the \emph{non-iid} nature of their datasets.
This defense essentially deduplicates at the model-update level rather than at the training-example level.%

Finally, in the extended version of this paper \cite[Appendix A]{debenedetti2023privacy}, we discuss another related poisoning defense, \emph{activation clustering} \citep{chen2018detecting} that is also vulnerable to similar side channels. %

\paragraph{The side channel}
The attacker introduces malicious samples so that the poisoning defense is triggered if and only if some target is present.
This yields a membership inference side channel.
Depending on the defense, this side channel can leak information about individual training samples or collections of samples (e.g., in the FL case when models updates represent information about a batch of examples).

\subsubsection{Our Attack}

For the semi-supervised poisoning defense of \citet{carlini2021poisoning}, the attack is exactly the same as for general data deduplication, so we do not discuss it further.
We thus focus on Sybil filtering in FL, which introduces some notable differences into the attack since the defense operates over model updates instead of individual samples.

We assume a threat model where the attacker controls one FL client and knows the defense mechanism being used.
The attacker's goal is to infer if some target client is participating in the protocol (i.e., client membership inference). We assume the attacker knows the target client's data distribution (e.g., if we are training a model on people's photos, then the attacker has some public photos of the target client). We do not assume that the attacker's data and the target's data overlap---only that they are similarly distributed.

We consider a canonical defense called FoolsGold \citep{fung2018mitigating}, which dynamically modifies the learning rate of individual clients based on the similarity of their updates over time. %
To compare the updates of two clients, the defense aggregates each client's history of updates and then computes the cosine similarity between these aggregates. A ``Sybil'' score between 0 and 1 is then computed for each client---based on the maximum similarity with another client---and this score is used to adjust the contribution of that client's model updates.

Our attack adds a malicious client into the training pool, with the aim of producing model updates that are similar to those produced by the target (if present).
The attacker then performs membership inference based on the model's loss on the \emph{attacker's} (not victim's) data:
if the target client is not participating in the protocol, FoolsGold will not downscale the attacker's updates and so the loss will decrease across epochs; conversely, if the target is present, the attacker's and target's updates are both downscaled and so the loss \emph{will not decrease}.
The attacker thus measures the model's loss on the attacker's dataset before and after participating in he FL protocol to determine if the target is present.

\subsubsection{Evaluation}

\begin{figure}[t]
    \centering
    \includegraphics[width=0.44\textwidth]{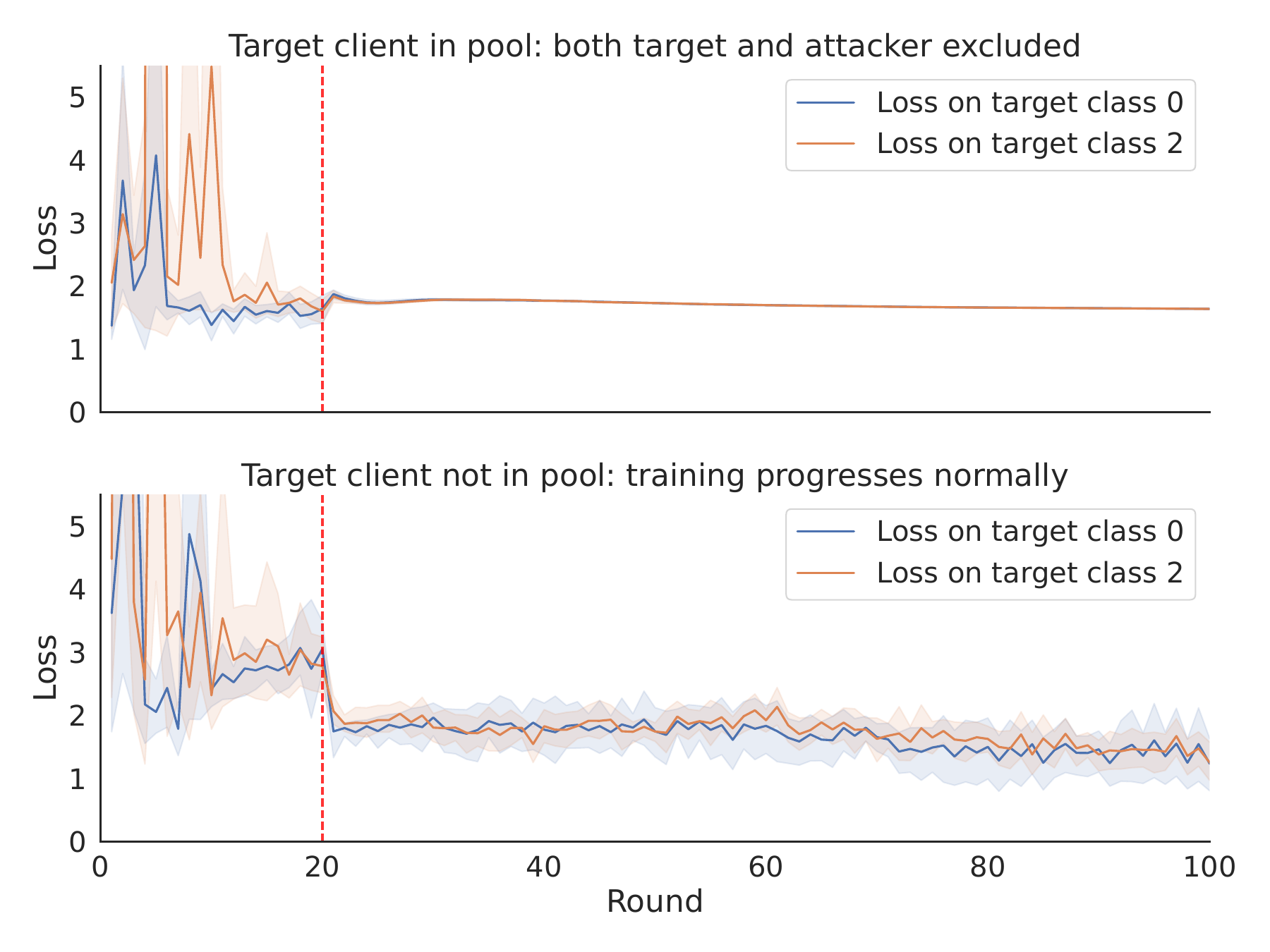}
    \vspace{-0.3cm}
    \caption{\textbf{When federated learning is combined with defenses against data poisoning, a side-channel is opened that worsens privacy.} We run the FoolsGold defense and insert a poisoning client into the learning protocol (denoted by the red vertical line). When a target client of interest is present in the data, both that client and the poisoning client will effectively not contribute to the learning (top of Figure). When the target is not present (bottom of Figure), the learning on the poisoning client continues as normal. This enables a strong membership inference attack.}
    \label{fig:fl_sybil_lfw}
    \vspace{-1.0em}
\end{figure}

We experiment with a non-iid setup that simulates a natural FL deployment to train an image classifier on pictures from each user's local device. We use the Labeled Faces in the Wild~\citep{LFWTech} (LFW) dataset, restricting the classification task to the 5 identities with over 100 pictures, and generate 5 clients each with 90 training images. To enforce the non-iid setup, we sample a majority (80\%) of images from a single identity for each client and then fill the remainder with images sampled from other classes according to a Dirichlet distribution ($\alpha = 1$). We train a ResNet-18~\citep{he2016deep} 
using the official FoolsGold implementation.\footnote{\url{https://github.com/DistributedML/FoolsGold}}
The malicious client joins the protocol at round 20; their dataset is composed of 50 non-overlapping images from the target client's identity. This ensures that its updates will be similar to the targeted client and different from others.
We run experiments with two target classes (0 and 2) and repeat each experiment 10 times with different random seeds.

\subsubsection{Results} 

\Cref{fig:fl_sybil_lfw} validates the susceptibility of FoolsGold to our side-channel attack.
As predicted, the model's loss on the attacker's dataset only decreases when the target is not present. 
By using the difference in average loss for 15 rounds before and after the attacker participated, we achieve 85\% accuracy (averaged over 10 experiments) in identifying whether the target client was in the training pool. 

Without the side channel, the adversary would need to make a membership guess based solely on the model's loss on the attacker's data.
A common approach from the literature would be to train shadow models. However, this is difficult in an FL setting as the attacker would need to know the data distribution of \emph{all} clients (not just the target), and perform multiple expensive simulations of the entire FL run.
We simulate such a strong adversary by taking the 10 different runs of our experiment, and setting the loss threshold directly as the difference between the model's average loss when the target is participating, or not.
We find that even such a strong adversary only achieves 67.5\% accuracy in guessing the presence of the target, without relying on our side channel.

\section{Attacking Input and Output Filters}
\label{sec:test_filters}

Here, we consider how filters applied to a model's inputs or outputs can enable powerful membership inference attacks. In fact, these membership inference attacks will be so strong (essentially perfect) that we can convert them into \emph{data extraction attacks} that even allow extraction of OpenSSH private keys contained in a language model's training set.

A common theme of this section is that ML system filters make it \emph{impossible} for the system to produce certain outputs, depending on the training data.
The attacker can then make some guess about an input in the training set, and craft a query that would trigger an impossible output if the guess were correct. This yields a perfect \emph{non-membership} inference attack: if the model produces the impossible output, the attacker knows their guess is incorrect. We then introduce various approaches to lift the attack into a strong (or even perfect) membership inference attack.
This attack strategy is reminiscent of impossible differential cryptanalysis~\citep{biham1999cryptanalysis, knudsen1998deal}, where an attacker's guess about the internal state of a cryptosystem can be ruled out if certain statistical output properties are observed.

\paragraph{Background: language models}
Large language models (LLMs) are neural networks that take as input a sequence of text of variable length and predict a probability distribution over the next word (or token, see below) in the sequence.
While early language models represented text as a sequence of characters or words~\citep{mikolov2010recurrent}, all recent language models use a more compact representation that splits arbitrary text into \emph{tokens} that represent entire words, sub-words, or characters.
The set of all tokens is called the \emph{vocabulary}.

When applied to an input sequence of tokens $t_1, t_2, \dots, t_n$, a LLM predicts a probability distribution 
$\Pr[t | t_1, t_2, \dots, t_n]$ for the value of the next token in the sequence.
To generate text, the model is applied repeatedly to its own output, using a particular \emph{decoding} strategy. In the simplest case of greedy decoding, we repeatedly sample the most likely token given the current input sequence and append this token to the input.

\subsection{Extracting Vocabularies of Language Models}
\label{ssec:vocab_extraction}

We first show how to use the input filtering stage of language models as a side channel.

\paragraph{Background: Byte-pair encoding} The inputs to modern language models are preprocessed into a series of sub-words (i.e., tokens) and truncated to a maximum context window size. The set of tokens (i.e., the model's vocabulary) is determined by running an algorithm such as byte pair encoding over the training data~\citep{sennrich2015neural}. \added{At the time of submission, Byte-pair encoding is the standard tokenization algorithm used in all modern large language models.}
A model's vocabulary can reveal sensitive details
of its training data, e.g., the GPT-2 tokenizer~\citep{radford2019language} contains tokens that represent individuals' Reddit usernames~\citep{rumbelow23solidgoldmagikarp}. \added{While this is not a privacy issue in the case of tokenizers trained on public data (like GPT-2's), this may become a concern if a tokenizer is trained or adapted on private data with an unusual vocabulary, e.g., medical records.}
Finally, many systems such as the Claude chatbot~\citep{claude} have their tokenizer kept secret.%

\paragraph{The side channel} Here, we show that truncating inputs to a fixed size window allows adversaries to extract the entire vocabulary of any model. 
In particular, consider a sequence such as {\color{red}``My favorite color is red. My favorite color is''}. Here, most language models will predict the word {\color{red}``red''} as the continuation. However, if some padding text of $\geq$ $N$ tokens is added, e.g., {\color{red}``My favorite color is red. {\color{blue}\texttt{PADDING}} My favorite color is''}, the model will not be able to see the word {\color{red}``red''} for long padding sequences.
This allows one to determine how many tokens the padding sequence occupies in the model's input.%

\subsubsection{Extracting the Complete Vocabulary} In particular,
under the byte-pair encoding algorithm, it is guaranteed that
any token can be recursively split into two sub-tokens until the final sub-tokens are only one byte.
Our attack will leverage this in reverse, where we initialize our extracted vocabulary $V$ to all of the single bytes and recursively expand it.
Concretely, for all pairs of tokens ${\color{blue}(u,v)} \in V \times V$,
we query the model on the sentence:
\[
\footnotesize
{\color{red}\textrm{``My favorite color is red.}}\  {\color{blue}\underbrace{u||v\ u||v\ \dots\ u||v}_{3N/4 \textrm{ times}}}\ {\color{red}\textrm{My favorite color is''}}
\]
where $||$ denotes concatenation. Suppose that ${\color{blue}u||v}$ actually \emph{is} a single token in the vocabulary.
Then the number of padding tokens inserted between the question and answer would be just $3N/4$ and the model would respond with ``red''.
In this case we insert the new token $V \gets V \cup \{{\color{blue}u||v}\}$.
On the other hand, if ${\color{blue}u||v}$ was represented instead by two tokens, then the number of padding tokens would be $3N/2 > N$ and the model will not answer ``red''.

\paragraph{Attack complexity} After $T$ iterations of this attack, we will recover all tokens of length less than $T$.
Extracting the entire vocabulary will require $O(|V|^T)$ total queries.

\paragraph{Results}
Our attack is empirically effective at recovering the entire GPT-2 vocabulary.
In the extended version of this paper~\cite[Appendix A]{debenedetti2023privacy}, Table 3 lists example of tokens that we extract, including rare substrings such as ``RandomRedditorWithNo'', ``TheNitromeFan'', and ``SolidGoldMagikarp''. Thes strings are usernames of Reddit users
that were likely repeated frequently in the GPT-2 training dataset.%
The attack requires $819{,}869{,}857$ queries. 

\subsubsection{Targeted Vocabulary Extraction}
We also can use this attack to efficiently extract the tokenization for a specific word or phrase.
Such an attack may be useful for (1) generating adversarial examples, where it may be useful to know the tokenization of a particular string, or (2) determining if a model uses another model's tokenizer, e.g., for licensing or reverse engineering purposes.

To use the above algorithm for a particular string such as ``hello'' and ``world'', we can simply check if ``he'', ``el'', ``ll'', etc. are tokens and would not need to check if ``ho'', ``wl'', ``ol'', etc. are tokens.%

\begin{figure}
    \centering
    \vspace{-1.5em}
    \includegraphics[scale=.75]{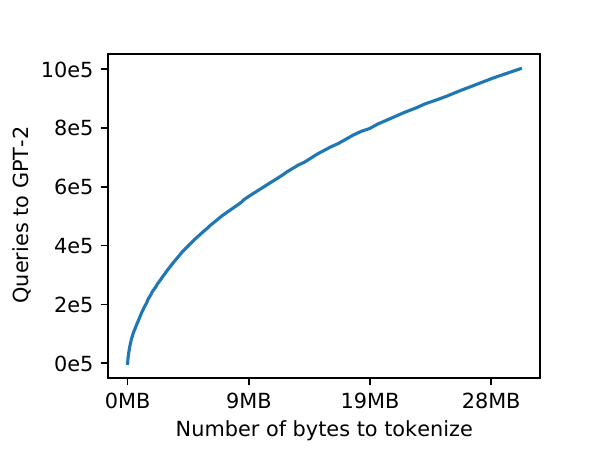}
    \vspace{-0.9em}
    \caption{\textbf{We extract the tokenizer for GPT-2 on specific byte strings from Wikipedia.} Our attack
    leverages a side channel based on the fact that language models use of a fixed context window.}
    \label{fig:enwik8}
    \vspace{-0.8em}
\end{figure}

\paragraph{Results}
In Figure~\ref{fig:enwik8} we measure the efficacy of this restricted attack by reporting
the number of queries necessary to determine how GPT-2 would tokenize
the first $N$ bytes of the enwiki8 dataset (a subset of Wikipedia).
On average we find it requires roughly 30,000 queries per megabyte,
diminishing as we increase the number of bytes to tokenize.

\subsection{Attacking Memorization Filters}
\label{ssec:memfree}

We next show how filters applied to LM outputs can also enable aggravated privacy violations.

\paragraph{Background: memorization-free decoding}
LLMs are known to memorize---and  output---sequences from their training dataset~\citep{carlini2021extracting}.
Thus, 
popular production LLM systems such as GitHub's Copilot
use filters that
block outputs that match their training data.
In this way, models are guaranteed to \emph{never} emit a verbatim sequence from their training dataset.

Ippolito \emph{et al.} \citep{ippolito2022preventing} formalize this as \emph{memorization-free decoding}.
Memorization-free decoding adds a filter that runs on-line with the language model and, before emitting the next token,
checks if this causes a k-gram match to a sequence in the training dataset (using an efficient Bloom filter lookup).
If so, this token is replaced with the most-likely non-memorized token instead.
While such filters are imperfect (for example they do not prevent the model from outputting similar yet not verbatim copies of training data), they are an efficient and practical defense that reduces the likelihood of data copying.

\paragraph{The side channel} We show that applying any form of memorization-free decoding introduces
a significant privacy vulnerability: if a language model 
\emph{ever} generates any particular k-gram, 
we are \emph{guaranteed} it was not part of the training data. 
This gives us a perfect \emph{non}-membership inference attack (i.e., a $100\%$ true negative rate at a 0\% false negative rate). Next, we will show how to convert this into a strong \emph{membership inference attack}.
\subsubsection{Our Attacks}\label{subsubsec:memfilter} We propose two different ways of extending the above non-membership inference attack to a near-perfect \emph{true positive rate}, thereby obtaining a near-perfect \emph{membership inference attack}.

\paragraph{Perfect membership inference for toggleable filters} To begin, we first develop a simple counterfactual-based approach for 
the special case where the memorization filter can be \emph{disabled}.
Indeed, some models, such as GitHub's Copilot model, allow the user to \textbf{choose} to block the recitation of training data.
This gives the adversary the ability to perform a perfect counterfactual analysis. %

Concretely, assume the adversary wants to perform (positive) membership inference on the sequence $t$.
We begin by \emph{disabling} the memorization filter and, for each
prefix $p_i$ of $t$, query the model on $p_i$ yielding a predicted suffix $s_i$.
For each of these, we check if $p_i || s_i$ is a prefix of $t$.

If none of the generations have this property then the attack is inconclusive.
However, if we identify \emph{any} index $i$ for which $p_i || s_i$ is a prefix of $t$, we can
\emph{enable} the memorization filter, and again prompt the model with $p_i$.
If the suffix with the filter turned on is equal to the original suffix $s_i$, then the sequence \emph{cannot} have been in the training dataset---if it was, the filter would have blocked it.
On the other hand, if the suffixes are not equal then we know with 100\% certainty that the sequence \emph{was} in the training dataset.

\paragraph{Membership inference with a permanent filter}
Some models contain memorization filters that are permanently applied, which prevents our straightforward counterfactual approach. The challenge with such a filter is that if a model fails to output some suffix, this could be due to one of two reasons: (1) either the filter was triggered (i.e., the sequence is in the training set); or the model simply assigns a low likelihood to this suffix.

To reduce the likelihood of the latter event, 
we propose to \emph{encourage} the model to emit the desired string, by exploiting the in-context learning abilities of language models.
For example, lets say we want to detect if the sequence {\color{blue}``ABCD''} is in the training set, and when we prompt the model with {\color{blue}``ABC''} it fails to output the letter {\color{blue}``D''}.
This is either because the memorization filter was triggered, or because the model simply assigns low probability to the completion {\color{blue}``D''} in this context.
To disambiguate these two cases, we prompt the model with the sequence {\color{blue}``ABCD ABCD ABCD ABC''}, where we repeat the targeted completion many times. Empirically, this guarantees that any powerful language model will output the completion {\color{blue}``D''} (as it is the most likely completion in the given context). Now, if the model \emph{still} fails to output {\color{blue}``D''}, we can assume with high confidence that the memorization filter has been triggered.
More formally, we split the target string
$t$ into a prefix $p_i$ and suffix $s_i$, prompt the model on
the string $p_i || s_i || p_i || s_i || \dots || p_i$, and then check if the output is $s_i$.

\subsubsection{Estimating Copilot's Training Date Cutoff}
We now describe an application of the toggleable filter side-channel attack to perform membership inference on GitHub Copilot, a popular public coding assistant. Concretely, Copilot uses a toggleable memorization filter that we exploit to determine the date on which GitHub scraped the model's training data.
The details of Copilot's memorization filter are not public, and in practice we found that it likely performs approximate filtering based on fuzzy matching between outputs and training data.
Nevertheless, we can run our attack as if it was an ideal filter
and accept that this could incur false positives
(approximate filtering does not cause false negatives).

\begin{figure}[t]
    \vspace{-1.5em}
    \centering
    \includegraphics[scale=0.4]{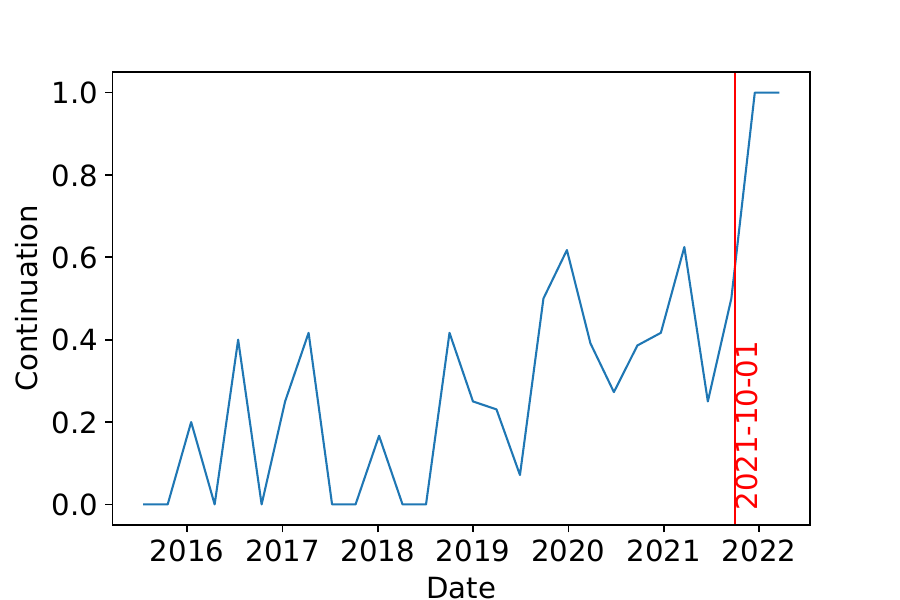}
    \vspace{-0.5em}
    \caption{\textbf{GitHub Copilot's memorization filter leaks membership information.} The system blocks outputs that are similar to the training data and thus reveals which code was used for training. In particular, we examine code commits that are added to a popular GitHub repository over time and plot the fraction of commits that our attack flags as non-members. There is a noticeable increase in October 2021, which reveals when the training data was last collected.}
    \label{fig:copilot_filter}
    \vspace{-1.9em}
\end{figure}

To evaluate our attack, we choose a popular GitHub repository, \texttt{tqdm}, and randomly sample some of the project's commits in which code was introduced that is still present
in the most recent revision of the project.
We then run our membership inference attack on each of these commits and in \Cref{fig:copilot_filter} we plot the fraction of commits our attack predicted were \textit{not} in the training set.

There are several false positives from before October 2021,
but for all commits made after that date, Copilot can generate the code with the filter on. This indicates that they are not members in the training data, and we can infer that October 2021 is likely the date at Copilot's data was last collected. This date is exactly the same date at which other OpenAI models' training data was collected (e.g., \texttt{text-davinci-003}), further confirming this result.

\added{Note that our attack shows that Copilot can generate code for all commits after October 2021 even while the filter is on. The attack does therefore not depend on the filter being toggleable. While in \Cref{subsubsec:memfilter} we give a baseline example of how an attack would work if the filter can be toggled, our full attack does not need this assumption. We choose \texttt{tqdm} as a target because it is a popular library with frequent commits, hence providing high resolution for our experiment. However, our attack still works with less popular repositories with less frequent commits. Finally, we attack Copilot's memorization filter as it is the only widely deployed filter we are aware of that matches the whole training corpus. Moreover, since the training data is already public, this experiment poses minimal risk. However, since we have shown a successful attack against a filter with unknown implementation details, we believe our attack should be generally applicable to other systems with a memorization filter.}

\subsubsection{From Output Filtering to Data Extraction}
Finally, we extend our attack to perform a data \emph{extraction} attack that can recover complete documents
from the training set token-by-token. We focus on the permanent memorization filter case, and we assume the attacker knows a substring of the text that they wish to extract from the training set.\footnote{We assume the substring is a prefix, but one can extend the attack to work with a known suffix by running membership inference attack on $s || p$.}
For example, to extract an RSA key the prefix could be ``BEGIN OPENSSH PRIVATE KEY''.

Given this prefix $p$ of a training document, a perfect membership inference attack makes it trivial to extend the sequence by one token:
enumerate all possible tokens $s$ that may occur next and for each token run the attack on $p || s$.
In practice, the attack may have false positives and thus identify several next-token candidates $s^*_j$. These false positives occur when the language model fails to emit a token that was not in the training set even when it is coerced to do so. 
In these cases, we explore each next-token candidate in a depth-first manner. If a candidate was incorrect, it will simply reach a state where there is no valid next token and terminate.

\paragraph{Extracting RSA private keys}
As a proof of concept, we consider the hypothetical case that GPT-Neo~\citep{gpt-neo} (a family of
language models similar to GPT-3) has a memorization filter and was trained on files containing some unknown RSA secret keys.
To instantiate the memorization filter, we build a Bloom filter containing all 20-token sequences in the GPT-Neo training
dataset and prevent the model from ever emitting these sequences.
To simulate training on private keys, we add $1{,}000$ different OpenSSH private keys to the Bloom filter, each of which are 512 base-64 encoded bytes.

Using this setup, we found that our membership inference attack achieves over $99.9\%$ accuracy at predicting whether a next-token candidate is in the training set.
However, this is still not enough for accurate data extraction as there are over $10{,}000$ possible next tokens in Base64 output and we want to extract text that contains many such tokens.%

To further enhance the attack, we count the frequency of each Base64-encoded token in a large 10TB dataset of random data. We then bias the search towards more common tokens, e.g., the Base64 token ``Q'' occurs much more frequently than ``omanip'' in practice (cfr Figure 9 in the extended version of this paper~\cite[Appendix A]{debenedetti2023privacy}).
Using this approach, we query the model with the complete header (``BEGIN OPENSSH PRIVATE KEY'') and run our attack. The attack extracts about 90\% of the keys successfully and requires around $340{,}000$ model queries (see \Cref{tab:rsaextract}).

\begin{table}[t]
\centering
\caption{\textbf{We can use an output filter side-channel to extract hundreds of secret OpenSSH private keys.} We use GPT-Neo language models with a permanent memorization filter and apply our membership inference approach to iteratively extract single tokens from the training set.}
\vspace{0.5em}
\begin{tabular}{@{}lrr@{}}
\toprule
Model & Success Rate & Mean Queries \\ \midrule
GPT-Neo 125M  & 90.3\%        & 376{,}000   \\
GPT-Neo 1.3B  & 90.0\%        & 338{,}000   \\
GPT-Neo 2.7B  & 89.8\%        & 344{,}000   \\ \bottomrule
\end{tabular}
\label{tab:rsaextract}
\end{table}

\section{Breaking Differentially Private Training}
\label{sec:dp}

We next show how the side channels that we have introduced thus far can violate ``provable'' privacy guarantees when those guarantees are based on isolated ML models.
Specifically, we show that standard differential privacy analysis fails when data filtering components are added.

\paragraph{Background: differentially private training}
An ML algorithm is differentially private (DP) if the distribution over possible trained models is close for any two neighboring training datasets that differ in a single example~\citep{dwork2006calibrating, abadi2016deep}. 
A common paradigm to design DP algorithms is to estimate the algorithm's \emph{sensitivity}---the maximum change in output from changing one input---and then adding random noise calibrated to that sensitivity.
However, estimating the sensitivity of a full ML training pipeline is hard.
Instead, the approach of the common algorithm DP-SGD~\citep{abadi2016deep} is to estimate the sensitivity of each training step and rely on DP's composition property to compute the model's final privacy budget $\epsilon$.

\paragraph{The side channel}
The theoretical analysis of DP-SGD does not consider additional components of the ML training pipeline such as filters applied before and after training. When these filters depend on the training data, they themselves have a non-zero sensitivity (i.e., changing one training example could change the filter's output). As a result, the total sensitivity of the system may be much larger than the sensitivity of the training algorithm in isolation.
This means that an ML system trained with DP-SGD could \emph{violate} the provable privacy guarantees that apply to an isolated model.

\paragraph{Warm-up: memorization filters have unbounded sensitivity}
As a simple example, consider the memorization-free decoding filter (\Cref{ssec:memfree}) that prevents the model from emitting training data. Its sensitivity is unbounded: the filter activates on a given output if and only if some training example matches it.
It is thus obvious that training the model with DP-SGD does not prevent the side-channel attacks that we describe in \Cref{ssec:memfree}.

\paragraph{Deduplication has large sensitivity}
A less obvious example is the case of data deduplication (\Cref{ssec:dedup}). One may expect that since deduplication empirically minimizes memorization, combining it with DP-SGD should lead to an extra layer of privacy protection. Indeed, prior work has even suggested this~\citep{ponomareva2022training}.
However, as our attacks in \Cref{ssec:dedup} show, deduplication can cause a single training example to influence whether many other examples get removed. In turn, the sensitivity of deduplication is quite large in practice.%

\paragraph{Evaluation}
We apply the side-channel attack from \Cref{ssec:dedup} to an ML system that first deduplicates the training set and then trains a model with DP-SGD.
We use the CIFAR-10 dataset and pick one training sample at random as the target. We create 256 approximate duplicates of the target and add these to the training set. We then deduplicate the training set using the default \texttt{imagededup} settings as in \Cref{ssec:dedup}.
We train 128 models using DP-SGD on either the full dataset as described above, or a neighboring dataset with the target removed.
Note that after deduplication, these initially-neighboring datasets now differ in 256 examples.
We use the state-of-the-art differentially-private training approach from~\citet{de2022unlocking}, with a batch size of 4096, augmentation multiplier of 16, and their settings for the noise multiplier and learning rate. 

We run DP-SGD multiple times with a target privacy budget $\epsilon \in (0.5,6)$ and set $\delta=10^{-5}$.
We then perform a membership inference attack on the adversarial duplicates (as  described in in \Cref{ssec:dedup}). We use the attack's FPR and TPR to estimate an empirical lower bound on the \emph{true} value of the privacy budget $\epsilon$ for the entire ML system---following prior work on DP auditing~\citep{nasr2021adversary,nasr2023tight}.

\begin{figure}[t]
    \centering
    \includegraphics[width=0.40\textwidth]{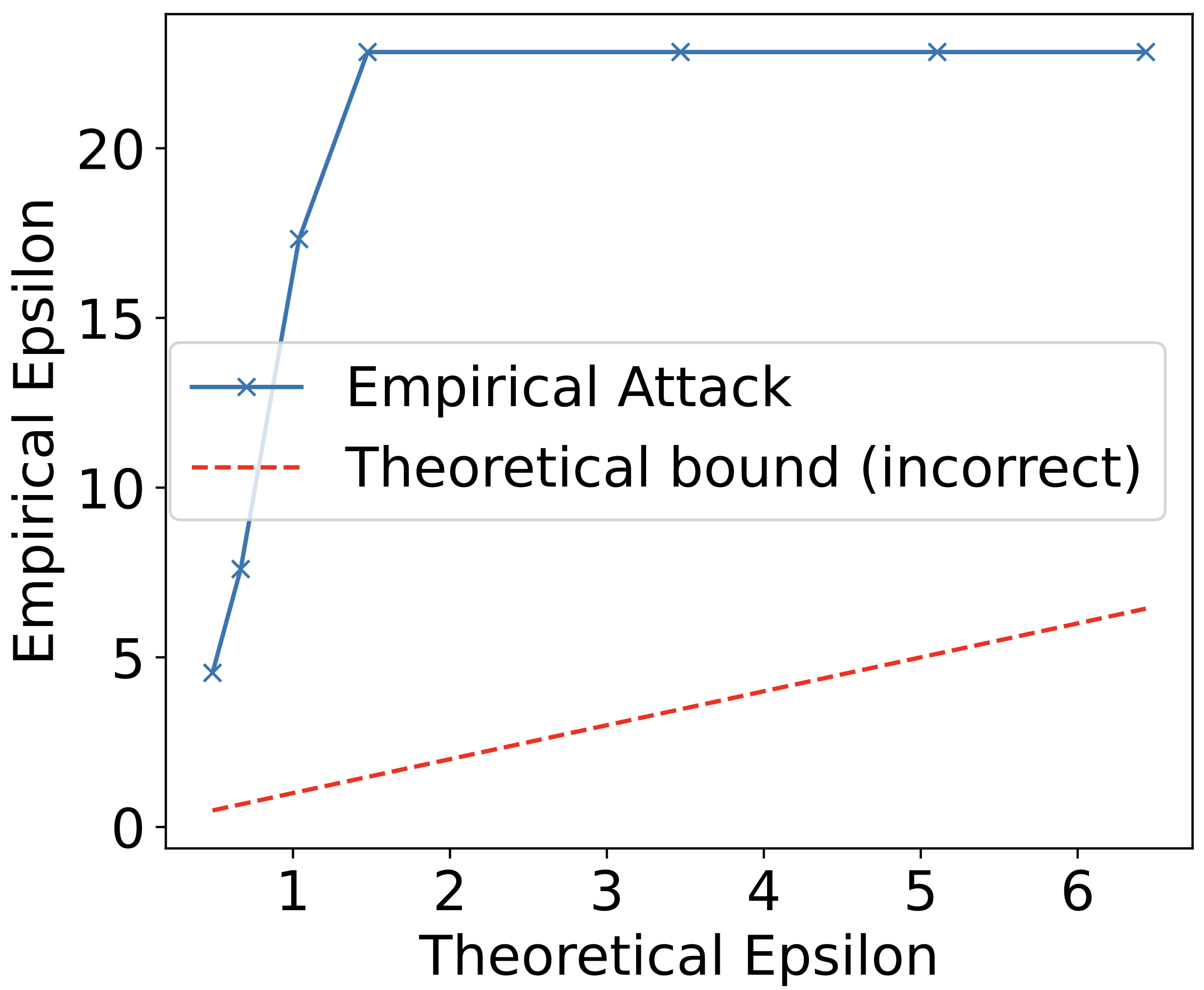}
    \vspace{-1em}
    \caption{\textbf{Deduplication violates naive differential privacy analysis.} If we combine approximate deduplication with DP-SGD (as suggested in prior work), then a single example in the collected training data can affect multiple samples in the  filtered training dataset that DP-SGD gets applied to. Deduplication thus effectively amplifies the sensitivity of the differentially private mechanism, leading to a severe underestimate of the true privacy loss of the end-to-end system.}
    \label{fig:dp_dedup}
    \vspace{-1em}
\end{figure}

\paragraph{Results} The privacy analysis of DP-SGD severely underestimates the privacy leakage of the entire system (Figure~\ref{fig:dp_dedup}). For example, when DP-SGD's theoretical bound on epsilon is respectively 1, 1.5, and 3, the actual privacy budget consumed by the system is \textit{at least} 7.5, 17, and 22.
Note this is not due to a mistake in DP-SGD's analysis or a bug in our implementation.
Rather, it is because the privacy analysis is with respect to the dataset that DP-SGD is actually run on and not the system's ``true'' training set $D_{\mathrm{train}}$ that is collected prior to deduplication. Thus, 
computing a correct privacy guarantee for the entire system would require composing DP guarantees for both the deduplication step and the training process.%

\section{Leaking Test Queries From Query Filters}
\label{sec:query_filters}

Thus far, we proposed side channel attacks that leak \textit{training} examples. Here, we show that side channels can also allow adversaries to identify \emph{test} queries made by arbitrary users. Worryingly, these attacks are otherwise impossible to launch when analyzing ML models in isolation.

Specifically, we focus on ML ``query filters'' that restrict or flag inputs in order to prevent attacks. These filters often guard against Sybil attacks (i.e., where an adversary uses multiple accounts to conceal their attack) by aggregating queries across \emph{all} users of the system. Here, we show that this introduces a side channel, where attackers can
create special queries whose outputs reveal information about other users' queries.

\subsection{Attacking Stateful Evasion Detectors}
\label{ssec:stateful_bbox_detectors}

\paragraph{Background}
In an evasion attack \citep{szegedy2013intriguing,biggio2013evasion}, an attacker perturbs an example
to cause a ML model to misclassify it.
To generate these adversarial examples in a black-box setting,
existing query-based attacks repeatedly query the target model~\citep{ilyas2018black}.
Given the difficulty in outright defending against evasion attacks,
recent work has developed \emph{stateful} detection defenses \citep{chen2020stateful,li2020blacklight}.
These defenses record every query (recently) issued to the system, and then ban a user (or take other defensive actions) if their query history appears malicious.
Commonly, a sequence of queries is marked as malicious if the queries are close to one another under some similarity metric.
As noted above, certain stateful evasion defenses also aggregate queries across users~\citep{li2020blacklight}.%

\paragraph{The side-channel}
We target the method of \citet{li2020blacklight}, \added{the first defense that proposes to store a global history of} fingerprint\added{s} (a fuzzy hash) of each query. If the current user's query is similar to a previous query, an attack is detected and the filter will reject the query and all subsequent similar queries. %
In turn, this enables a side channel where one can check if another user has uploaded some query $x$ by re-uploading the same query. If the system rejects the attacker's query, they know that a query very similar to $x$ was likely previously uploaded. %

\paragraph{Evaluation and results} We evaluate our attack on CIFAR-10. We create a set $D_\text{test}$ of $1{,}000$ images to simulate queries made by users and a held-out set $D_{\sim \text{test}}$ of $1{,}000$ other images.
We query the model on each image in $D_\text{test}$ and $D_{\sim \text{test}}$ and predict which images were real queries (i.e., $D_\text{test}$).
Our attack has perfect accuracy (100\% TPR, 0\% FPR): the defense rejects the attacker's query if and only if the image was previously queried.
The only possible failure case for the attack is a false positive, which the defense is explicitly set up to minimize and is thus rare in practice.
Overall, this side channel highlights the strong tension between security and privacy---if one uses a privacy-preserving defense that acts on a per-user basis, they cannot resist Sybil attacks.\footnote{Note that it may be possible to improve the defense. For example, instead of filtering a query when it is similar to \emph{one} prior query, the defender could pick a random threshold $T$ on the number of similar past queries. We leave an exploration of this to future work.}

\subsection{Extension to Stateful Extraction Detectors}
\label{ssec:stateful_extraction_detectors}

Aside from black-box evasion attacks, stateful detectors also exist for model extraction attacks~\citep{tramer2016stealing, papernot2017practical}. In extraction attacks, adversaries make many queries to a black-box ML model in order to reverse-engineer a local copy with similar functionality. 
Existing stateful defenses against model extraction~\citep{liu2022seinspect, pal2021stateful, zhang2021seat, juuti2019prada} 
keep track of queries and search for examples that are highly similar to one another or are consecutively out-of-distribution.
To date, no extraction methods aggregate across users and are thus all vulnerable to Sybil attacks. However, if they were generalized to use a global query log (as suggested in \citep{li2020blacklight} for the PRADA defense in \citep{juuti2019prada}), then they would become vulnerable to a similar privacy side channel as we described in Section~\ref{ssec:stateful_bbox_detectors}.

\subsection{Extension to Retrieval-based Text Detection}

Finally, similar side channel attacks exist for retrieval-based text detectors~\citep{krishna2023paraphrasing}. Here, 
the goal is to test whether some generation $y$ was produced by a particular language model. \citet{krishna2023paraphrasing} approach this by recording each output from a language model and then checking whether any text similar to $y$ was previously generated by the model. Unfortunately, such methods allow adversaries to check if some output was produced as a response to another user's query, as acknowledged by \citep{krishna2023paraphrasing}.

\section{Conclusion}
\label{sec:conclusion}

In this work, we showed that real-world ML systems can have drastically worse privacy than is otherwise suggested by typical standalone privacy analyses. In particular, we showed how adaptive adversaries can use black-box or white-box system knowledge to exploit side channels in popular ML components. Taken together, these attacks contribute to a rethinking of how practitioners should measure and mitigate privacy risks for state-of-the-art ML systems.

Perhaps most surprisingly, it is actually the introduction of components that are \textit{intended to improve privacy} (i.e., data deduplication and memorization-free decoding) that leads to side-channels that cause privacy violations. Similarly, methods designed to protect against other forms of adversarial attacks (e.g., data poisoning and evasion attacks) can also backfire and induce privacy violations at elevated rates.
These findings highlight the inherent tensions between security and privacy, and how improving privacy in the average case can have adverse effects when faced with worst-case adversaries.

Moving forward, there are numerous rich areas for future work in the space of system-level analyses of ML systems. In particular, while we focus specifically on privacy side-channels, our work hints at the possibility that other side-channels may exist for different types of attacks and threat models. Moreover, our work highlights open problems in ensuring privacy-preserving ML, such as how to enforce differential privacy when composing hard-to-analyze ML modules such as filters or anomaly detectors. We hope to tackle these and other challenges in future work.

\section*{Author Contributions}
\begin{itemize}[noitemsep]
\item Nicholas and Florian proposed the problem statement of side-channel leakage in ML system components.
\item Florian proposed the side channels in data deduplication and stateful detectors.
\item Florian and Matthew proposed the side channels in data poisoning defenses.
\item Nicholas and Eric proposed the side channel in memorization filters.
\item Nicholas proposed the side channel to extract vocabularies of LLMs.
\item Edoardo performed experiments on data deduplication and stateful evasion detectors, and drafted the corresponding sections of the paper.
\item Giorgio and Matthew performed experiments on data poisoning, and drafted the corresponding sections.
\item Nicholas performed experiments on memorization filters and extracting vocabularies, and drafted the corresponding section of the paper.
\item Milad performed experiments with GitHub Copilot and on attacking DP-SGD, and drafted the corresponding sections of the paper.
\item Edoardo, Giorgio, Nicholas, Milad, and Eric designed the paper's figures.
\item Edoardo, Giorgio, Nicholas, Christopher, Matthew, Milad, Eric, and Florian wrote the paper.
\item Florian organized the project.
\end{itemize}

\section*{Acknowledgments}
We thank Reza Shokri, Jamie Hayes, Andreas Terzis, Vijay Bolina, Helen King, Eve Novakovic, Amanda Carl, Jenna LaPlante, Sanah Choudry, and Daniel Paleka for helpful discussions. We also thank the reviewers and the shepherd for the useful feedback.
Eric Wallace is supported by the Apple Scholars in
AI/ML Fellowship. Edoardo Debenedetti is supported by armasuisse Science and Technology.

\bibliography{references}
\bibliographystyle{plainnat}

\appendix

\section{Additional Experiments and Figures}

\subsection{Data Deduplication}
\label{apx:dedup-backdoor} 
\begin{figure}[h]
    \centering
    \includegraphics[width=0.4\textwidth]{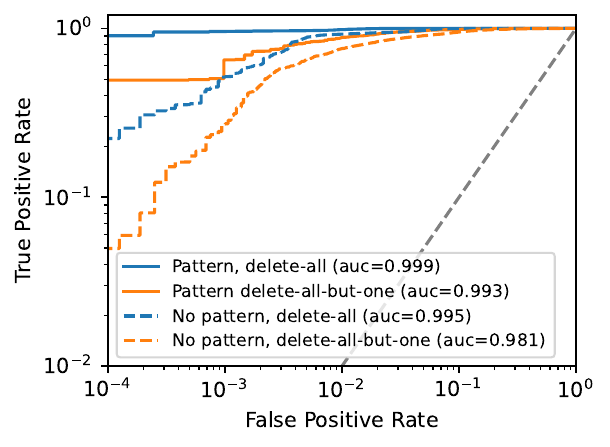}
    \vspace{-0.3cm}
    \caption{\textbf{Using a backdoor-like pattern improves membership inference.} We show the effectiveness of our attack with 8 approximate duplicate images, both with and without a backdoor pattern. In either the \emph{delete-all} or \emph{delete all-but-one} settings, the pattern increases privacy leakage.}
    \label{fig:backdoor}
\end{figure}

\cref{fig:backdoor} shows that using a backdoor pattern significantly strengthens membership inference. The backdoor ensures that the model confidently memorizes all the attacker's near-duplicates, when they are not deduplicated.

\subsection{Additional Side Channels in Poisoning Defenses}
\label{app:pois}
Activation clustering \citep{chen2018detecting} is a defense against backdoor attacks. The defender first clusters the training data based on the similarity of their late-layer representations (computed after training an initial model). Then, clusters that are smaller than some minimal threshold $T$ are removed. This defense relies on two assumptions: (1) poisoned examples are likely to be similar to each other in representation space (and dissimilar from natural data) and (2) the adversary cannot add $T$ or more poisoning examples.

\paragraph{The side channel}
The side channel for activation clustering is similar to that for deduplication. In order to target an example $(x, y)$, the adversary can inject $T-1$ copies of that example. These will necessarily be clustered together, such that, if the original target example is present, the cluster will contain at $T$ copies of that example, and if the target example is not present, the cluster will contain $T-1$ copies of that example. Then activation clustering will always remove the cluster when the target is present, but will not remove the cluster if the target is not present, unless other examples are clustered together with the copies. This allows for a near-perfect non-membership inference (the $T-1$ copies will lead to very low loss on the example) but imperfect membership inference (other examples in the cluster will lead to false positives).

\paragraph{Evaluation}
We implement activation clustering on the CIFAR-10 dataset, using $k$-means clustering with $k=600$ clusters and a threshold of $T=11$. 
Setting $k=600$ reflects a setting where the defense is designed to catch much smaller attacks than existing implementations. We train 50 ResNet-9 models on random 50\% splits of CIFAR-10 and target 50 distinct examples by adding 10 copies of that example to all models. 
When a cluster has 11 or more examples, we label it as containing the target, and when the cluster has only 10 examples, we label it as not containing the target. 
As predicted, we observe perfect non-membership inference with this information: all clusters with $<11$ examples cannot contain the target. However, we observe a large false positive rate and have an overall membership inference precision of only 70.0\% due to the presence of other benign examples in the target clusters. Thus, a privacy side channel exists but its effectiveness depends on the specific hyperparameters and data distribution.

\subsection{Exploiting Tokenizers}

\begin{figure}[h]
    \centering
    \includegraphics[scale=.7]{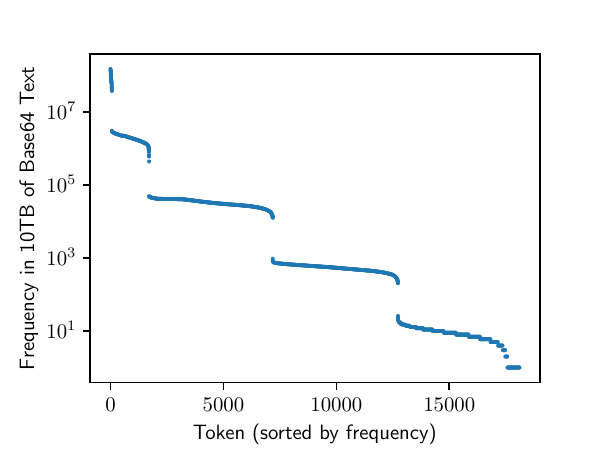}
    \vspace{-0.2cm}
    \caption{\textbf{Some GPT-2 tokens are $\mathbf{10{,}000{,}000\times}$ more frequent than others in Base-64 encoded text.}
    This lets us perform an efficient greedy search to extract RSA keys by
    minimizing the number of backtracks required.}
    \label{fig:tokenfreq}
\end{figure}

\Cref{tab:gpt2tokens} gives results for the vocabulary extraction attack on GPT-2 from \Cref{ssec:vocab_extraction}.
\Cref{fig:tokenfreq} plots the frequency of each Base-64 token, for the attack on memorization
filters from \Cref{ssec:memfree}.

\begin{table}[h]
    \centering
    \caption{\textbf{We extract the complete tokenizer of GPT-2 using our side-channel.}
    We report the number of LM queries required to recover all tokens of each length, as well as the number of tokens of that length in the vocabulary. We also show one example token for each length.%
    }
    \vspace{0.5em}
    \label{tab:gpt2tokens}
    \resizebox{\columnwidth}{!}{\begin{tabular}{@{}rrrl}%
    \toprule
         Length & Query Count & Unique Tokens & Example Token \\ %
         \midrule
2 & 35{,}532 & 1767 & ĵĺ \\
3 & 647{,}079 & 5006 & âĵĺ \\
4 & 5{,}026{,}516 & 7161 & iHUD \\
5 & 21{,}385{,}519 & 7290 & srfN \\
6 & 59{,}335{,}183 & 6415 & UCHIJ \\
7 & 121{,}674{,}066 & 5912 & attRot \\
8 & 203{,}113{,}848 & 5053 & Adinida \\
9 & 295{,}349{,}381 & 3841 & quickShip \\
10 & 390{,}944{,}926 & 2811 & DevOnline \\
11 & 482{,}954{,}060 & 1924 & StreamerBot \\
12 & 565{,}643{,}235 & 1126 & GoldMagikarp \\
13 & 635{,}651{,}132 & 612 & PsyNetMessage \\
14 & 691{,}713{,}142 & 336 & TheNitromeFan \\
15 & 734{,}125{,}866 & 129 & RandomRedditor \\
16 & 764{,}520{,}511 & 56 & InstoreAndOnline \\
17 & 785{,}211{,}929 & 16 & natureconservancy \\
18 & 798{,}586{,}326 & 11 & SolidGoldMagikarp \\
19 & 806{,}854{,}922 & 5 & guiActiveUnfocused \\
20 & 811{,}773{,}192 & 1 & \texttt{--------------------} \\
21 & 814{,}622{,}879 & 3 & RandomRedditorWithNo \\
         \bottomrule
    \end{tabular}}
\end{table}

\end{document}